\documentclass[prx,aps,twocolumn, amssymb,amsmath,floatfix,superscriptaddress]{revtex4-2}

\usepackage{graphicx}
\usepackage{dcolumn}
\usepackage{bm}% bold math
\usepackage{hyperref}
\usepackage{amsthm}

\usepackage{physics} 
\usepackage{comment}
\usepackage{lineno}
\usepackage{braket}
\usepackage{tabularx}
\usepackage{bm}
\usepackage{euscript}
\usepackage{epsfig,psfrag}
\usepackage{amsthm}
\theoremstyle{definition}

\usepackage{graphicx}
\usepackage{color}
\usepackage{amsfonts}
\usepackage{exscale}
\usepackage{wrapfig}
\usepackage{amsthm}

\usepackage{extarrows} 
\usepackage{placeins}
\usepackage{soul}
\usepackage{amsmath}
\usepackage[english]{babel}
\usepackage[autostyle, english = american]{csquotes}
\MakeOuterQuote{"}
\usepackage{enumitem}
\usepackage{slashed} 
\usepackage{tikz}

\usepackage{bm}
\usepackage{braket}
\usepackage{enumitem}
\usepackage{xcolor}
\usepackage{hyperref}
\hypersetup{colorlinks=true,linkcolor=blue!55!black,citecolor=blue!55!black,urlcolor=blue!55!black}

\numberwithin{equation}{section} 
\renewcommand{\theequation}{\arabic{section}.\arabic{equation}} 
\newcommand{\bea}{\begin{eqnarray}}  
\newcommand{\eea}{\end{eqnarray}}
\newcommand{\ben}{\begin{enumerate}}
\newcommand{\een}{\end{enumerate}}
\newcommand{\be}{\begin{equation}}
\newcommand{\ee}{\end{equation}}

\definecolor{Bcolor}{RGB}{0,0,255}
\definecolor{Rcolor}{RGB}{255,0,0}
\definecolor{Gcolor}{RGB}{0,255,0}

\usetikzlibrary{external}
\newcommand{\wloss}{\omega_{\mathrm{loss}}}
\newcommand{\win}{\omega_{\mathrm{in}}}
\newcommand{\wout}{\omega_{\mathrm{out}}}

\newcommand{\dg}{^{\dagger}}

\begin{document}

\title{Generating entanglement in quantum matter with pathway-entangled photons via resonant inelastic X-ray scattering}

\author{Chunjing Jia}
\email{chunjing@phys.ufl.edu}
\affiliation{Department of Physics, University of Florida, Gainesville, Florida 32611, USA}

\author{Valentin Walther}
\affiliation{Department of Chemistry, Purdue University, West Lafayette, Indiana 47907, USA}
\affiliation{Department of Physics and Astronomy, Purdue University, West Lafayette, Indiana 47907, USA}

\author{Laimei Nie}
\email{nlm@purdue.edu}
\affiliation{Department of Physics and Astronomy, Purdue University, West Lafayette, Indiana 47907, USA}

\date{\today}

\begin{abstract}
Resonant inelastic X-ray scattering (RIXS) is conventionally used to probe elementary excitations  and correlations in quantum materials. Here, we propose an interferometric RIXS protocol that instead creates entanglement between spatially separated regions of a solid-state material. An incident X-ray photon is coherently split by a beam splitter into two arms illuminating opposite ends of a sample. Recombining the scattered photons at a second beam splitter and detecting an energy-selected photon then projects the material into a coherent superposition of excitations at the two ends. This protocol is the X-ray, solid-state analogue of the Duan-Lukin-Cirac-Zoller (DLCZ) heralding scheme. Using a Kramers-Heisenberg formalism with quantized photon modes, we show that orthogonal and degenerate end excitations yield maximally entangled states. We illustrate this mechanism in three models: (1) two decoupled sites yield a heralded Bell state. (2) An open tight-binding chain exhibits reflection-parity-dependent interference, providing a benchmark for the interferometer's coherence. (3) A spin-1 Haldane (AKLT) chain yields an entangled pair of fractionalized edge spins. We also discuss practical considerations and challenges for realizing the protocol in X-ray experiments.
\end{abstract}

\maketitle

% =============================================================================
\section{Introduction}
\label{sec:intro}
% =============================================================================

Entanglement is the defining nonclassical resource of quantum mechanics~\cite{Horodecki2009} and the organizing principle behind topological order, symmetry-protected phases and the efficiency of tensor-network descriptions~\cite{Amico2008,Laflorencie2016}. In synthetic quantum matter including ultracold-atom and trapped-ion simulators, microscopic control and site-resolved measurements have enabled the preparation and characterization of entangled many-body states \cite{Islam2015,Brydges2019}. However, achieving comparable access to entanglement in solid state materials remains challenging. Important progress has come from entanglement witnesses based on quantum Fisher information, connecting many-body entanglement to response functions accessible through inelastic neutron scattering \cite{Hauke2016,Scheie2021,Laurell2021} and resonant inelastic X-ray scattering (RIXS) \cite{Hales2023,Ren2024,TenHuisen2026}. Another development is the preparation of neutrons with entangled spin, path, and energy degrees of freedom \cite{Shen2020}, together with theoretical proposals to use such probes to access entanglement in matter~\cite{Irfan2021}.

These advances establish scattering as a means of characterizing entanglement. A distinct possibility is to use the detection of scattered radiation itself to prepare an entangled state between spatially separated degrees of freedom, %Photon-mediated heralding implements this idea through interference and conditional detection
as proposed by Cabrillo et al. and Duan, Lukin, Cirac, and Zoller \cite{Cabrillo1999,DLCZ2001}. Related protocols have generated entanglement between distant nitrogen-vacancy-center spins \cite{Bernien2013}, spins in separate semiconductor quantum dots \cite{Delteil2016,Stockill2017}, and vibrational excitations in separate diamond crystals \cite{Lee2011}. These demonstrations motivate the extension of photon-mediated heralding to spatially separated degrees of freedom within a single correlated material.

Here, we propose such a route using resonant inelastic X-ray scattering (RIXS) with path-entangled X-ray photons. RIXS is a powerful spectroscopic probe of collective excitations in correlated materials, with applications ranging from spin and orbital excitations to charge and lattice dynamics.~\cite{Ament2011, schlappa2012spin, devereaux2016directly, jia2014persistent, hepting2020electronic,rossi2019experimental} In parallel, {\it X-ray quantum optics} has advanced through photon-correlation and parametric-conversion experiments~\cite{Adams2013,Shwartz2014, Volkovich2020, Hartley2025}. More recently, Powers and Durbin demonstrated Hong–Ou–Mandel-type two-photon interference using synchrotron X-rays~\cite{PowersDurbin2026}. These developments represent important progress in X-ray interferometric techniques.

\begin{figure}[t]
\includegraphics[width=\columnwidth]{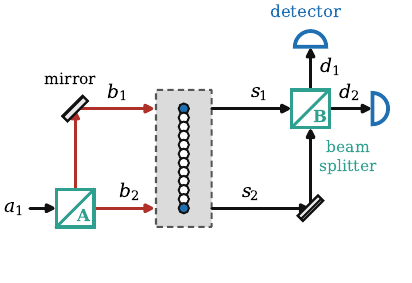}
\caption{\label{fig:setup}
Pathway-entangled-photon RIXS on a one-dimensional sample. The incident mode $a_1$ is coherently divided by a $50/50$ beam splitter $A$ into arms $b_1$ and $b_2$, which are focused on the two ends $j=1$ and $j=N$ of the chain (gray box; the two illuminated end sites are shown in blue). The scattered modes $s_1,s_2$ are recombined on a second $50/50$ splitter $B$ and registered by detectors $d_1,d_2$. In the energy-resolved configuration analyzed in the main text, an energy analyzer in each arm (not drawn) transmits only the inelastically scattered photon at $\hbar\wout=\hbar\win-\hbar\wloss$, so that a single click at $d_1$ or $d_2$ heralds the prepared state. }
\end{figure}

To combine RIXS with photon-mediated heralding, we consider a Mach-Zehnder interferometer (Fig.~\ref{fig:setup}): a beam splitter $A$ divides the incident photons into arms $b_1$ and $b_2$ that illuminate two distinct locations on the sample.  
The scattered beams $s_1,s_2$ are then recombined at a second $50/50$ splitter $B$ and detected at $d_1,d_2$. 
When scattering at the two locations creates orthogonal but degenerate material excitations, recombination at $B$ erases information about where the RIXS scattering occurred. A detector click then projects the sample into a coherent superposition of the two scattering outcomes, entangling the two regions. 

To illustrate this mechanism, we first consider an idealized setting with a single photon prepared in a superposition of the two incident arms. An energy analyzer in each arm selects the inelastically scattered photon at $\hbar\omega_{\rm out}=\hbar\omega_{\rm in}-\hbar\omega_{\rm loss}$, allowing a detector click to herald the material state. This is the direct X-ray analogue of the DLCZ protocol~\cite{DLCZ2001}. Our main results are:

\begin{enumerate}
\item General criterion (Sec. II). A detector click at $d_1$ or $d_2$ prepares a coherent superposition $|\mu_1\rangle\pm e^{i\Phi}|\mu_N\rangle$, where $|\mu_1\rangle$ and $|\mu_N\rangle$ are the material states produced by scattering at the two ends.

\item Two-site toy model (Sec.~\ref{sec:twosite}). The limit of $\braket{\mu_1|\mu_N}=0$ is achieved and a single click heralds the Bell state $\ket{eg}\pm e^{i\Phi}\ket{ge}$, where $\ket{g}$ is the local ground state and $\ket{e}$ is the state after RIXS. % with unit fidelity, and the count rate is flat.

\item Tight-binding chain with open boundary condition (Sec.~\ref{sec:tb}). The opposite limit is realized, where scattering at both ends produce the same delocalized particle–hole excitation, so no entanglement is generated. Nevertheless, this can be used as a calibration tool: the reflection parity of the final states determines which detector receives the photon, providing a direct test of the interferometer's coherence and phase stability (Fig.~\ref{fig:tb}).

\item The spin-1 Haldane chain (Sec.~\ref{sec:haldane}) as a more realistic model. The localized spin-$1/2$ edge modes provide nearly degenerate excitations at opposite ends. A detector click heralds an entangled state of the two edge spins, increasing the middle-cut entanglement entropy by $\ln 2$ in our AKLT chain calculation. 

\end{enumerate}

%We then extend our analysis to more realistic experimental settings, including phase averaging associated with imperfect beam splitters, as motivated by the Powers-Durbin experiment, and chaotic (thermal) light sources characteristic of synchrotron radiation. We briefly discuss how these effects influence the heralding protocol and under what conditions entanglement can still be generated. We also discuss a complementary photon number-resolved detection scheme that does not require energy resolution (Appendix A).
We also discuss an alterative two-photon detection scheme in Appendix~\ref{app:counting} and conclude by outlining the unresolved challenges of experimental implementation and possible directions for further investigation.

% =============================================================================
\section{Formalism}
\label{sec:formalism}
% =============================================================================

\subsection{Interferometer and mode conventions}
\label{sec:modes}

The two $50/50$ splitters act on photon creation operators as
\begin{equation}
a_{b_1}\dg=\tfrac{1}{\sqrt2}\big(a_{a_1}\dg + a_{a_2}\dg\big),\qquad
a_{b_2}\dg=\tfrac{1}{\sqrt2}\big(a_{a_1}\dg- a_{a_2}\dg\big),
\label{eq:splitterA}
\end{equation}
at $A$, and
\begin{equation}
a_{s_1}\dg=\tfrac{1}{\sqrt2}\big(a_{d_1}\dg + a_{d_2}\dg\big),\qquad
a_{s_2}\dg=\tfrac{1}{\sqrt2}\big(a_{d_1}\dg- a_{d_2}\dg\big),
\label{eq:splitterB}
\end{equation}
at $B$. A single photon  arriving at beam splitter $A$ will undergo 
\be
|1,0 \rangle_a \to \tfrac{1}{\sqrt2}\Big[ |1,0 \rangle_b + |0,1\rangle_b \Big].
\ee
where the subscripts $a, b$ label the pathways. Similarly, a photon state $\alpha\ket{1,0}_s+\beta\ket{0,1}_s$ arriving at
$B$ wil be mapped to
\begin{equation}
\alpha\ket{1,0}_s+\beta\ket{0,1}_s \to \tfrac{1}{\sqrt2}\Big[(\alpha+\beta)\ket{1,0}_d+(\alpha-\beta)\ket{0,1}_d\Big].
\label{eq:onephoton}
\end{equation}

\subsection{Kramers--Heisenberg amplitude with quantized photon modes}
\label{sec:KH}

The RIXS intensity in the standard formulation is
\begin{equation}
\mathcal{I}(\win,\wloss)=\sum_F\Big|\sum_I
\frac{\bra{F}D'\ket{I}\bra{I}D\ket{G}}{E_G-E_I+i\Gamma}\Big|^2
\delta\big(E_F-E_G\big),
\label{eq:KH}
\end{equation}
where $\ket{G},\ket{I},\ket{F}$ are the initial, intermediate, and final joint states of electron and photon with corresponding energies $E_G, E_I, E_F$, $\Gamma$ is the inverse core-hole lifetime, and $D, D'$ are the dipole operators of the photon absorbing and emitting processes,
\begin{align}
D &=\lambda_k \sum_{j=1}^N e^{i\bm{k}\cdot\bm{r}_j}\,
    c_{c,j}\dg c_{o,j}\,a_{k\epsilon,j},
\label{eq:D}\\
D'&=\lambda_{k'} \sum_{j = 1}^N e^{-i\bm{k}'\cdot\bm{r}_j}\,
    c_{o,j}\dg c_{v,j}\,a\dg_{k'\epsilon',j}.
\label{eq:Dp}
\end{align}
Here $c_o, c_v, c_c$ are the electron annihilation operators at the core, valence, and conduction bands respectively. $a_{k \varepsilon}$ ($a\dg_{k \varepsilon}$) is the photon annihilation (creation) operator for momentum $\bm{k}$ and polarization $\epsilon$. $\lambda_{k}$ ($\lambda_{k'}$) denote the light-matter coupling strengths and contain the corresponding dipole matrix elements.
It is convenient to work with the scattered state,
\begin{equation}
\ket{\Psi}\equiv D'\,\mathcal{G}\,D\ket{G},
\qquad
\mathcal{G} \equiv \frac{1}{E_G-H_0+i\Gamma},
\label{eq:Psi}
\end{equation}
where $H_0$ is the bare Hamiltonian of the electron and photon system without the electron-photon coupling part. For a single-photon input, the initial state is $|G\rangle=|g\rangle\otimes|1_{\bm{k}\varepsilon}\rangle$. Projecting the scattered state in Eq.~\eqref{eq:Psi} onto a final electron state $|f\rangle$ and an outgoing photon state $|1_{\bm{k}'\varepsilon'}\rangle$, and factoring out the light-matter coupling coefficients and spatial phases, we obtain the local RIXS amplitude
\begin{equation}
M_j(f)=\Bra{f}\,c_{v,j}\,
\frac{1}{\hbar\win-(H_{\mathrm{int}}-E_g)+i\Gamma}\,c_{c,j}\dg\,\Ket{g},
\label{eq:Mj}
\end{equation}
where $H_{\mathrm{int}}$ is the electron Hamiltonian of the intermediate state with the presence of the core hole.

\subsection{Entanglement heralding}
\label{sec:heralding}

Now we apply the knowledge above to the setup in Fig~\ref{fig:setup}. The dipole operators are defined on the two ends,
\begin{align}
D &=\lambda_k \! \! \sum_{j \in \{1, N \}} \! \! e^{i\bm{k}\cdot\bm{r}_j}\,
    c_{c,j}\dg c_{o,j}\,a_{k\epsilon,j},
\\
D' &=\lambda_{k'}\!\! \sum_{j \in \{1, N \}} \!\! e^{-i\bm{k}'\cdot\bm{r}_j}\,
    c_{o,j}\dg c_{v,j}\,a\dg_{k'\epsilon',j}.
\end{align}
Let each arm carry an analyzer between the sample and beam splitter $B$ that transmits photons within a narrow energy window around $\hbar\wout=\hbar\win-\hbar\wloss$ and rejects photons at $\hbar\win$. The state reaching splitter $B$ is then
\begin{equation}
\begin{aligned}
\ket{\Psi}\propto\sum_f\Big[&e^{i\phi_1}M_1(f)\,\ket{f}\otimes\ket{1,0}_s\\ +&e^{i\phi_N}M_N(f)\,\ket{f}\otimes\ket{0,1}_s\Big],
\end{aligned}
\label{eq:psi_c}
\end{equation}
where the sum runs over material final states $|f\rangle$ whose corresponding scattered photon energies fall within the analyzer window, and
\begin{equation}
\phi_j \equiv (\bm{k}-\bm{k}')\cdot\bm{r}_j.
\label{eq:phase}
\end{equation}
Applying Eq.~\eqref{eq:onephoton} and dropping the global phase $e^{i\phi_1}$, we obtain the state reaching the detector after beam splitter $B$:
\begin{equation}
\boxed{
\begin{aligned}
\ket{\Psi}\propto\ &\big[\,\ket{\mu_1}+e^{i\Phi}\ket{\mu_N}\,\big]\otimes\ket{1,0}_d\\ +\,&\big[\,\ket{\mu_1}-e^{i\Phi}\ket{\mu_N}\,\big]\otimes\ket{0,1}_d ,
\end{aligned}}
\label{eq:heralded}
\end{equation}
where we have defined
\be
\Phi = \phi_N-\phi_1,
\ee
\begin{equation}
\ket{\mu_1}=\sum_f M_1(f)\ket{f},\qquad
\ket{\mu_N}=\sum_f M_N(f)\ket{f}.
\label{eq:mu}
\end{equation}
Therefore, a click at $d_1$ heralds $\ket{\mu_1}+e^{i\Phi}\ket{\mu_N}$, and a click at $d_2$ heralds $\ket{\mu_1}-e^{i\Phi}\ket{\mu_N}$. 
The corresponding count rates are
\begin{equation}
\begin{aligned}
I_{10(01)}=\tfrac12\sum_f\big|M_1(f)\pm e^{i\Phi}&M_N(f)\big|^2\\[-2pt]
&\times\delta(\hbar\wloss-E_f+E_g)
\end{aligned}
\label{eq:I1001}
\end{equation}
Their sum
\begin{equation}
I_{10}+I_{01}=\sum_f\big(|M_1(f)|^2+|M_N(f)|^2\big)\,\delta(\hbar\wloss-E_f+E_g)
\label{eq:sumrule}
\end{equation}
is independent of the splitter. This is because $B$ is unitary and preserves the total scattered intensity. It is also independent of the relative phase $\Phi$, which only affects how the intensity is distributed between the two detectors. All of the interferometric information sits in the difference,
\begin{equation}
I_{10}-I_{01}=2\sum_f\mathrm{Re}\big[e^{i\Phi}M_1^{*}(f)M_N(f)\big]\,
\delta(\hbar\wloss-E_f+E_g).
\label{eq:difference}
\end{equation}

\noindent Two limiting cases are particularly instructive:
\paragraph*{(i) $\braket{\mu_1|\mu_N}=0$.}
If scattering at the two ends produces orthogonal and degenerate final states $\ket{f_L}$ and $\ket{f_R}$, the corresponding material states become $\ket{\mu_1}=M\ket{f_L}$ and $\ket{\mu_N}=M\ket{f_R}$, where $M\equiv M_1(f_L)=M_N(f_R)$. The heralded states
\begin{equation}
\ket{\Psi_\pm}=\tfrac{1}{\sqrt2}\big(\ket{f_L}\pm e^{i\Phi}\ket{f_R}\big)
\label{eq:bell}
\end{equation}
are maximally entangled across a cut separating the two ends, with entanglement
entropy $\ln 2$.
The count rates are $I_{10} = I_{01} = \frac{1}{2} |M|^2$. This is the two-site case of
Sec.~\ref{sec:twosite} and the Haldane chain in Sec.~\ref{sec:haldane}.

\paragraph*{(ii) $\ket{\mu_1}\parallel\ket{\mu_N}$.}
If both ends feed the same final state $\ket{\mu_N}=e^{i\theta}\ket{\mu_1}$, then the heralded state $\propto(1\pm e^{i(\Phi+\theta)})\ket{\mu_1}$ is proportional to $\ket{\mu_1}$, and $\max_{\Phi}(I_{10} - I_{01}) / (I_{10} + I_{01}) = 1$. 
In this limit, the relative phase $\theta$ determines how the scattered photons are distributed between the two detectors, and no entanglement is generated in the material. This is the tight-binding case of Sec.~\ref{sec:tb}.

The discussion above focuses on a single-photon input. It is worth noting that two-photon inputs also provide a route to heralded entanglement. For example, with one photon incident on the sample on each arm, $|1,1\rangle_b$, joint detection of the outgoing photons can also prepare entangled material states, as shown in Appendix~\ref{app:counting}. Such protocols may offer greater robustness against certain optical phase fluctuations, an aspect we will investigate systematically in a forthcoming work.

% =============================================================================
\section{Two decoupled sites: a heralded Bell state}
\label{sec:twosite}
% =============================================================================

\begin{figure}[t]
\includegraphics[width=\columnwidth]{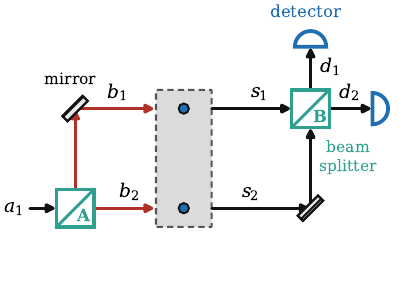}
\caption{\label{fig:twosite}
The two-site limit of Fig.~\ref{fig:setup}. The sample is reduced to two decoupled sites, each placed under an optical pathway. An excitation created at site $1$ cannot propagate, so the two ends feed the orthogonal, degenerate final states $\ket{eg}$ and $\ket{ge}$. This is the solid-state counterpart of the two atomic ensembles of the DLCZ protocol \cite{DLCZ2001}, with the two illuminated sites playing the role of the ensembles and the core-hole resonance playing the role of the Raman drive.}
\end{figure}

Let us first consider a toy model of two identical and decoupled sites, one per arm (Fig.~\ref{fig:twosite}). Each site contains a core, valence, and conduction level, with energies $\varepsilon_o,\varepsilon_v, \varepsilon_c$ respectively. We denote the local ground state by $\ket{g}$ where the core and valence levels are occupied, and the excited state by $\ket{e}$ where the core and conduction levels are occupied after RIXS.
Their energy difference is $\varepsilon_c - \varepsilon_v$. Starting from $\ket{gg}$, RIXS at site $1$ and $2$ produce $\ket{eg}$ and $\ket{ge}$, respectively. Because there is no hopping, the local amplitudes Eq.~\eqref{eq:Mj} are diagonal,
\begin{equation}
M_1(f)=M\,\delta_{f,eg},\qquad M_N(f)=M\,\delta_{f,ge},
\label{eq:Mtwosite}
\end{equation}
with $M \propto 1/(\hbar\win-(\varepsilon_c - \varepsilon_o)+i\Gamma)$. Equation~\eqref{eq:heralded} becomes
\begin{equation}
\boxed{
\begin{aligned}
\ket{\Psi}\propto\ &\big(\ket{eg}+e^{i\Phi}\ket{ge}\big)\otimes\ket{1,0}_d\\
             +\,&\big(\ket{eg}-e^{i\Phi}\ket{ge}\big)\otimes\ket{0,1}_d ,
\end{aligned}}
\label{eq:twosite_state}
\end{equation}
and the count rates are
\begin{equation}
I_{10}=I_{01}=\tfrac12|M|^2\,\delta(\hbar\wloss-(\varepsilon_c- \varepsilon_v)).
\label{eq:twosite_I}
\end{equation}
Since $|eg\rangle$ and $|ge\rangle$ are degenerate, the outgoing photon energy does not reveal which site underwent RIXS. A single click at $d_1$ prepares the triplet-like combination $(|eg\rangle+e^{i\Phi}|ge\rangle)/\sqrt{2}$, while a click at $d_2$ prepares $(|eg\rangle-e^{i\Phi}|ge\rangle)/\sqrt{2}$. Both heralded states are maximally entangled with a $\ln 2$ single-site entanglement entropy.

% =============================================================================
\section{Tight-binding chain: benchmark and parity readout}
\label{sec:tb}
% =============================================================================

We now consider the opposite limit where the excitation created at either end is delocalized over the whole sample. Consider a single band on an open chain of $N$ sites,
\begin{equation}
H=(\varepsilon_0-\mu)\sum_j c_j\dg c_j-t\sum_j\big(c_j\dg c_{j+1}+\mathrm{h.c.}\big),
\label{eq:Htb}
\end{equation}
whose eigenstates are standing waves
\begin{equation}
\psi_n(j)=\sqrt{\tfrac{2}{N+1}}\,\sin\tfrac{n\pi j}{N+1}, \quad n = 1, ..., N
\label{eq:tbstates}
\end{equation}
with energies
\begin{equation}
\varepsilon_n=\varepsilon_0 - \mu -2t\cos\tfrac{n\pi}{N+1}.
\label{eq:tbenergies}
\end{equation}
Unlike the two-site case, the occupied and unoccupied orbitals here belong to the same band, while the localized core levels involved in RIXS are treated implicitly.
We take the initial state $\ket{g}$ to be the ground state at filling $N_e/N=1/2$, with the lowest $N_e$ single-particle orbitals occupied. During RIXS, a core electron at energy $\varepsilon_o$ is first excited into orbital $p$ at energy $\varepsilon_p$, leaving a core hole, which is subsequently filled by an electron from orbital $q$ at $\varepsilon_q$. Denoting $\gamma_n^\dagger=\sum_j\psi_n(j)c_j^\dagger$ as the electron creation operator on the $n$th orbital, the final states after RIXS contain a particle–hole pair $\ket{p\bar q}=\gamma_p^\dagger\gamma_q\ket{g}$, where $q\leq N_e$ labels an initially occupied orbital and $p>N_e$ an initially unoccupied orbital. We further set the ground state energy to zero, so the final electronic state has energy $E_{p\bar q}=\varepsilon_p-\varepsilon_q$.

Neglecting the core-hole potential, Eq.~\eqref{eq:Mj} becomes
\begin{equation}
M_j(p\bar q)=\frac{\psi_q(j)\,\psi_p(j)}{\hbar\win-(\varepsilon_p + \mu - \varepsilon_o)+i\Gamma}.
\label{eq:Mtb}
\end{equation}
up to an overall prefactor. 
%Reflection about the chain center relates the two end amplitudes. 
Since $\psi_n(N)=(-1)^{n+1}\psi_n(1)$,
\begin{equation}
M_N(p\bar q)=(-1)^{p+q}\,M_1(p\bar q)%\equiv P_f\,M_1(p\bar q),
\label{eq:parity}
\end{equation}
When the energy analyzer resolves a particular particle–hole excitation $\ket{p\bar q}$, scattering at either end produces the {\it same} delocalized final state, with amplitudes differing only by a sign determined by the reflection parity of the excitation. Therefore, a detector click leaves the material in $\ket{p\bar q}$, creating no entanglement.

\begin{figure*}[t]
\includegraphics[width=\textwidth]{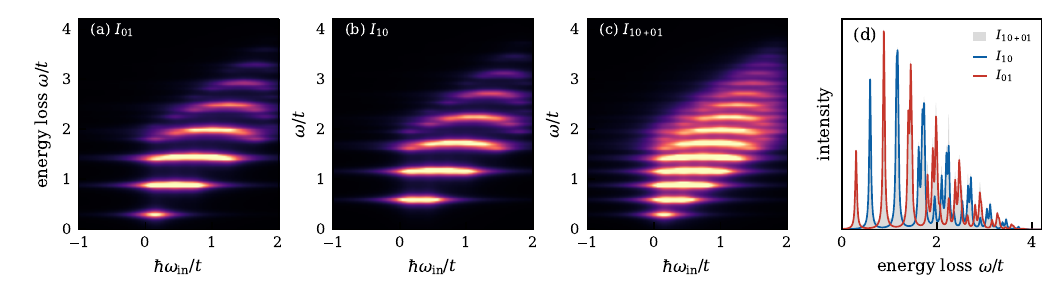}
\caption{\label{fig:tb}
Entangled-photon RIXS spectra of the open tight-binding chain Eq.~\eqref{eq:Htb}, at $\Phi=0$, $N=20$ sites, $\Gamma/t=0.20$, and half filling. Here $\omega = \wloss$. (a)(b) Energy-loss maps versus incident energy $\hbar\win/t$ for the two detector channels. By Eq.~\eqref{eq:parity} the two channels collect mutually exclusive sets of final states, sorted by the reflection parity $(-1)^{p+q}$, i.e. every line present in (a) is absent in (b) and vice versa. (c) Their sum, which is independent of the beam splitter and of $\Phi$ based on the sum rule Eq.~\eqref{eq:sumrule}. 
(d) Energy-loss cut comparing the two parity-resolved spectra against the total spectrum (gray).} 
\end{figure*}

However, the same parity dependence provides a useful way to test the interferometer's coherence and phase stability. For a scattering geometry with $\Phi = (\bm{k}-\bm{k}')\cdot (\bm{r}_N - \bm{r}_1) = 0$, substituting Eq.\eqref{eq:parity} into Eq.~\eqref{eq:I1001} gives
\begin{equation}
\begin{aligned}
I_{10}&\propto\!\!\sum_{p+q\ \mathrm{even}}\!\!|M_1(p\bar q)|^2\,\delta(\cdots),\\
I_{01}&\propto\!\!\sum_{p+q\ \mathrm{odd}}\!\!|M_1(p\bar q)|^2\,\delta(\cdots),
\end{aligned}
\label{eq:parityfilter}
\end{equation}
i.e., the interferometer acts as a perfect parity filter, with every reflection-even final state routed deterministically to one detector, every reflection-odd final state to the other. This property follows from reflection symmetry and is independent of the Hamiltonian details. Any leakage into the opposite detector would indicate imperfect interference between the two scattering paths. 

Figure~\ref{fig:tb}  illustrates this parity-dependent routing. Panels (a) and (b) show the detector count rates $I_{10}$ and $I_{01}$, respectively, where each particle–hole excitation contributes to only one detector according to its reflection parity. Panel (c) shows their sum, $I_{10}+I_{01}$, which reproduces the total RIXS intensity without beam splitter $B$, consistent with Eq.~\eqref{eq:sumrule}. Panel (d) compares the two detector spectra at a fixed incident photon energy, highlighting the separation of even- and odd-parity contributions.

% =============================================================================
\section{Haldane chain: heralded entanglement of fractionalized edge spins}
\label{sec:haldane}
% =============================================================================

The two-site and tight-binding models illustrate the key requirements for generating entanglement between the two ends of a sample: RIXS at the two ends must produce localized and degenerate excitations. We now turn to a more physically relevant interacting system: a spin-1 chain in the Haldane phase. Its fractionalized spin-$1/2$ edge modes naturally provide such excitations, which are nearly degenerate and separated from bulk excitations by the Haldane gap $\Delta_H$~\cite{Haldane1983PLA,Haldane1983PRL}. 

To illustrate this physics, we consider the Affleck-Kennedy-Lieb-Tasaki (AKLT) model~\cite{AKLT1987,AKLT1988},
\begin{equation}
H=J\sum_i\Big[\bm{S}_i\cdot\bm{S}_{i+1}
+\tfrac13\big(\bm{S}_i\cdot\bm{S}_{i+1}\big)^2\Big],
\label{eq:aklt}
\end{equation}
Its ground state is a valence-bond solid, where each spin-1 is represented by two spin-$1/2$ degrees of freedom, with neighboring sites connected by singlets. An open chain has two unpaired spin-$1/2$ states at the ends, forming effective edge spins $s_L$ and $s_R$.

\subsection{Spin-flip RIXS and heralded Bell states}

Potential experimental realizations include Ni-based spin-1 Haldane chain materials such as NENP and Y$_2$BaNiO$_5$~\cite{Renard1987,Hagiwara1990,NENP2025, Y2BaNiO5}. At the Ni $L_{2,3}$ edge, spin–orbit coupling of the $2p$ core hole enables a spin-flip RIXS channel~\cite{Ament2011, jia2014persistent}. More precisely, in a cross-polarization channel, the effective local RIXS operator $D'_j\mathcal{G}D_j$ is proportional to the spin-lowering operator $S_j^-$.~\cite{jia2014persistent} 
Furthermore, a small magnetic field can polarize the edge states into $\ket{\uparrow_L\uparrow_R}$ and give their spin-flip excitations a finite energy $g\mu_B B$. %This can separate the edge excitations from elastic scattering and the bulk continuum. 

\begin{figure*}[t]
\includegraphics[width=\textwidth]{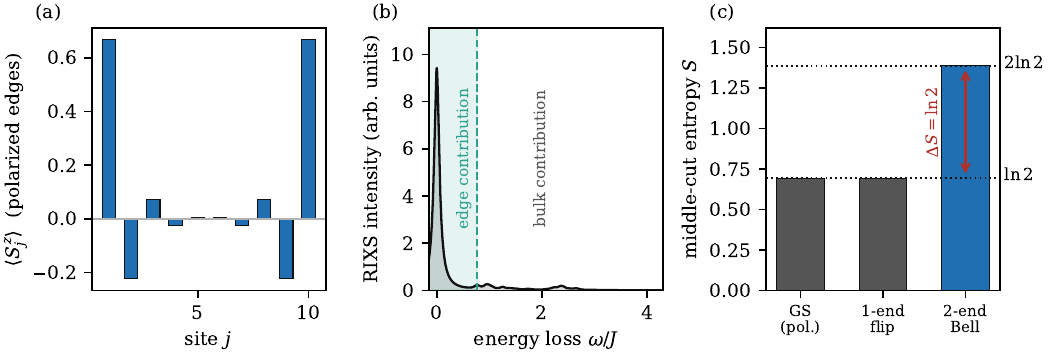}
\caption{\label{fig:haldane}
Entangling the two ends of an AKLT/Haldane chain: exact diagonalization of Eq.~\eqref{eq:aklt} with $N = 10$ sites. (a) Local magnetization $\langle S^z_j\rangle$ of the ground state with polarized edges, showing the exponentially localized edge modes. (b) RIXS response  $\sum_f|\bra{f}S^-_1\ket{g}|^2\delta(\hbar\wloss-E_f+E_g)$ at the end site $j=1$, with $\omega = \wloss$. A sharp in-gap edge peak (shaded) carries most of the end-flip weight, well separated from the bulk magnon continuum above the Haldane gap $\Delta_H$ (dashed line). The analyzer window of Sec.~\ref{sec:heralding} is set on this peak. (c) Entanglement entropy of half chain. Both the ground state with polarized edge spins and any single-end flip on top of the ground state produced by conventional one-beam RIXS give $S=\ln 2$, which comes from the  contribution of the middle bond. The heralded two-end Bell state of Eq.~\eqref{eq:haldane_state} gives $S=2\ln 2$.}
\end{figure*}
With $\ket{\mu_1} \propto \ket{\downarrow_L\uparrow_R}$ and $\ket{\mu_N} \propto \ket{\uparrow_L\downarrow_R}$ which are orthogonal and degenerate, Eq.~\eqref{eq:heralded} becomes
\begin{equation}
\boxed{
\begin{aligned}
\ket{\Psi}\propto\
&\big(\ket{\downarrow_L\uparrow_R}+e^{i\Phi}\ket{\uparrow_L\downarrow_R}\big)
 \otimes\ket{1,0}_d\\
+\,&\big(\ket{\downarrow_L\uparrow_R}-e^{i\Phi}\ket{\uparrow_L\downarrow_R}\big)
 \otimes\ket{0,1}_d ,
\end{aligned}}
\label{eq:haldane_state}
\end{equation}
so that a click at $d_1$ or $d_2$ heralds a maximally entangled state of the two edge spins, with entanglement entropy $\ln 2$. 
Furthermore, similar to the case of two decoupled sites,  the count rates are again independent of $\Phi$. %, $I_{10}=I_{01} \propto \delta(\hbar\wloss-g\mu_B B)$. 
In the calculation below, we set the initial state to be the polarized state $\ket{\uparrow_L\uparrow_R}$, and set external field $B \rightarrow 0$.

\subsection{Exact diagonalization}

Here we calculated two different RIXS processes for the AKLT chain: the conventional RIXS process with the excitation only at single edge (site 1); our proposed pathway-entangled-photon RIXS process as shown in Fig.~\ref{fig:setup}.  The exact diagonalization results of a 10-site AKLT chain are summarized in Fig.~\ref{fig:haldane}. Panel (a) shows the local magnetization $\langle S^z_j\rangle$ of the polarized ground state, which is large at each end and exponentially small in the bulk. The conventional spin-flip RIXS response at site 1 has the expression 
\begin{equation}
\begin{aligned}
&\sum_f|M_1(f)|^2\,\delta(\hbar\wloss-E_f+E_g)
\end{aligned}
\end{equation}
, where $M_1(f) = \bra{f}S^-_1\ket{g}$. The calculated result for such quantity as shown in Panel (b) reveals a sharp low-energy peak, separated from bulk excitations by the Haldane gap $\Delta_H$. Such energy loss profile stays the same when pathway-entangled-photon RIXS is implemented.
Panel (c) further compares the entanglement entropy across the middle of the AKLT chain for three different states: the ground state, the final state obtained by single end spin flip RIXS process, the final state obtained by pathway-entangled-photon RIXS process. Both the ground state and the state obtained by flipping a single edge spin using conventional RIXS have $S=\ln 2$, arising from the valence bond crossing the middle cut. In contrast, heralding a Bell state of the two edge spins using our protocol increases the entropy to $2\ln 2$. The increase of the entanglement entropy shows a clear picture that our proposed protocol generates entanglement in the AKLT chain.

% =============================================================================
\section{Summary and outlook}
\label{sec:summary}
% =============================================================================

We have developed a theoretical framework for generating entanglement in quantum materials through interferometric RIXS. Assuming coherence between the two interferometer arms and ideal energy selection, detecting a scattered photon prepares a superposition of material excitations created at two spatially separated locations. We illustrate this mechanism through Bell-state preparation in a two-site model and entanglement generation between the fractionalized edge spins of an AKLT chain. The tight-binding example provides a complementary benchmark, where interference sorts the final states according to their reflection parity. These results establish a conceptual basis for using interferometric RIXS prepare quantum states.

%These theoretical results open many opportunities for using entangled X-ray photons in conventional X-ray science. A key question is how to design new classes of experiments in which entangled X-rays interact with materials in qualitatively new ways, both to generate entanglement in matter and to witness the resulting entangled states. Another important direction is to extend the scope beyond RIXS toward broader applications in X-ray science, including X-ray diffraction, X-ray absorption spectroscopy, and related techniques.

%In realistic experiments, realizing the proposed protocol also presents many challenges. A central experimental challenge is overcoming the intrinsically weak RIXS signal and the associated limitations of existing X-ray instrumentation, while still preserving and detecting the relevant quantum entanglement.
Translating this mechanism into an experiment presents substantial challenges. First, realistic X-ray sources exhibit photon-number fluctuations rather than producing Fock states. Our preliminary analysis suggests that such fluctuations alone do not necessarily destroy heralded entanglement. Second, imperfections in the optical elements can introduce uncontrolled relative phase fluctuations between the two interferometer arms. A promising solution to this is two-photon input and joint detection. A third challenge is the limited energy resolution of existing X-ray analyzers, which may not be sufficient to distinguish the desired low-energy spin-flip excitations from elastic scattering and bulk excitations in the Haldane chain example. Biphoton-selective diffraction exploiting the half-wavelength interference effect~\cite{Shih2001}, combined with alternative interferometer designs, could potentially suppress unwanted photon backgrounds and reduce the demands on energy resolution. Finally, verifying the generated entanglement and determining its lifetime may be possible through a pump–probe experiment, where a second X-ray pulse could reveal characteristic interference signatures distinguishing the prepared entangled state from the initial unentangled state, or provide an independent entanglement witness through quantum Fisher information extracted from RIXS measurements~\cite{Hales2023,Ren2024,TenHuisen2026}. A detailed investigation of some of these effects will be presented in a forthcoming work.

A perhaps even bigger challenge is the following: unlike conventional RIXS spectroscopy where weak inelastic signals can be identified by accumulating sufficient statistics, entanglement heralding requires distinguishing the desired scattering events from elastic scattering and other unwanted processes. %Maintaining coherence between the two interferometer arms and verifying the resulting material entanglement are also experimentally demanding. 
%Whether these challenges can be addressed simultaneously in a realistic experiment in the short term remains an open question.
Addressing these challenges simultaneously in a realistic experiment remains an important longer-term goal.
Meanwhile, these challenges motivate the exploration of simpler interactions between pathway-entangled X-rays and matter, such as scattering processes with stronger signals that do not require energy-loss resolution. More broadly, our results raise the question of how the quantum properties of X-ray photons could be exploited to generate or probe entanglement in materials. Extending these ideas beyond RIXS to techniques such as X-ray diffraction and X-ray absorption spectroscopy may open new directions in X-ray quantum science.

% =============================================================================
\begin{acknowledgments}
We are grateful to Stephen Durbin, Liam Powers, Eric Schultz, Erez Berg, Wei-Sheng Lee, Yuan-Ron Shen, Zhurun (Judy) Ji, Wanli Yang,
Yujian Xia, Jiarui Li, Steve Kivelson, and Lingjia Shen for illuminating discussions and valuable comments on this work. ChatGPT and Claude were used to assist the generalization from two-site toy model to tight binding chain as well as the exact diagonalization of the AKLT chain. VW acknowledges support from the U.S. Department of Energy (DOE), Office of Science, Basic Energy Sciences (BES), under Award $\#$DE-SC0026318, and by the National Science Foundation (NSF) under Award $\#$PHYS-2409630.
\end{acknowledgments}
% =============================================================================

\appendix
\renewcommand{\theequation}{A\arabic{section}.\arabic{equation}}
% =============================================================================
\section{Photon-number-resolved detection with an indistinguishable pair}
\label{app:counting}
% =============================================================================

The energy-resolved protocol of Sec.~\ref{sec:heralding} achieves the erasure of path information via a frequency filter. An alternative is to obtain it from photon statistics, at the cost of requiring an indistinguishable  photon pair onto the sample. We lay out the procedure below. 

Take the input to be $\ket{G}=\ket{gg}\otimes\ket{1,1}_b$, one photon in each arm, the two photons being indistinguishable. To leading order exactly one photon is absorbed and re-emitted: either the arm-1 photon scatters, in which case arm 2 still carries its $\hbar\win$ photon, or vice versa. The scattered state $|\Psi \rangle$ in Eq.~\eqref{eq:Psi} is then
\begin{equation}
\begin{aligned}
\ket{\Psi}=\mathcal{N}\sum_f\Big[&e^{i\phi_1}M_1(f)\,\ket{f}\otimes\ket{1_{k'},1_k}_s\\
+&e^{i\phi_N}M_N(f)\,\ket{f}\otimes\ket{1_k,1_{k'}}_s\Big],
\end{aligned}
\label{eq:psi11}
\end{equation}
where we have specified the momenta  of the input ($k$) and outgoing ($k'$) photons.
Recombining on beam splitter $B$ using Eq.~\eqref{eq:splitterB} sorts this into the two measurable photon-number patterns,
\begin{equation}
\begin{aligned}
\ket{\Psi}\propto\sum_f\Big[
&\big(e^{i\phi_1}M_1+e^{i\phi_N}M_N\big)\ket{f}\otimes\ket{2002}_d\\
+&\big(e^{i\phi_1}M_1-e^{i\phi_N}M_N\big)\ket{f}\otimes\ket{1111}_d\Big],
\end{aligned}
\label{eq:psi_sorted}
\end{equation}
with the shorthand
\begin{align}
\ket{2002}_d&\equiv\ket{2_{kk'},0}-\ket{0,2_{kk'}} &&\text{(no coincidence)},\\
\ket{1111}_d&\equiv\ket{1_k,1_{k'}}-\ket{1_{k'},1_k} &&\text{(coincidence)}.
\end{align}
The first pattern sends both photons to the same detector (photon bunching), while the second places one photon at each detector. Projecting and squaring gives the energy-loss-resolved intensities
\begin{equation}
\begin{aligned}
I_{2002/1111}=\tfrac12\sum_f\big|e^{i\phi_1}M_1(f)&\pm e^{i\phi_N}M_N(f)\big|^2\\[-2pt]
&\times\delta(\hbar\wloss-E_f+E_g),
\end{aligned}
\label{eq:I2002}
\end{equation}
which is formally identical to the energy-resolved result Eq.~\eqref{eq:I1001} under $I_{2002}\to I_{10}$, $I_{1111}\to I_{01}$. This means Fig.~\ref{fig:tb} also holds for the photon-counting case.

\bibliographystyle{apsrev4-2}
\bibliography{refs}

@article{Ament2011,
  title = {Resonant inelastic x-ray scattering studies of elementary excitations},
  author = {Ament, Luuk J. P. and van Veenendaal, Michel and Devereaux, Thomas P. and Hill, John P. and van den Brink, Jeroen},
  journal = {Rev. Mod. Phys.},
  volume = {83},
  issue = {2},
  pages = {705--767},
  numpages = {0},
  year = {2011},
  month = {Jun},
  publisher = {American Physical Society},
  doi = {10.1103/RevModPhys.83.705},
  url = {https://link.aps.org/doi/10.1103/RevModPhys.83.705}
}

@article{Adams2013,
  author  = {Adams, B. W. and Buth, C. and Cavaletto, S. M. and Evers, J. and Harman, Z. and Keitel, C. H. and P{\'a}lffy, A. and Picon, A. and R{\"o}hlsberger, R. and Rostovtsev, Y. and Tamasaku, K.},
  title   = {X-ray quantum optics},
  journal = {J. Mod. Opt.},
  volume  = {60},
  number  = {1},
  pages   = {2--21},
  year    = {2013},
  doi     = {10.1080/09500340.2012.752113}
}

@article{Shwartz2014,
  title = {X-Ray Parametric Down-Conversion in the Langevin Regime},
  author = {Shwartz, S. and Coffee, R. N. and Feldkamp, J. M. and Feng, Y. and Hastings, J. B. and Yin, G. Y. and Harris, S. E.},
  journal = {Phys. Rev. Lett.},
  volume = {109},
  issue = {1},
  pages = {013602},
  numpages = {5},
  year = {2012},
  month = {Jul},
  publisher = {American Physical Society},
  doi = {10.1103/PhysRevLett.109.013602},
  url = {https://link.aps.org/doi/10.1103/PhysRevLett.109.013602}
}

@article{Hartley2025,
  author  = {Hartley, N. J. and others},
  title   = {X-ray parametric down-conversion at an {XFEL}},
  journal = {Optica},
  volume  = {12},
  number  = {7},
  pages   = {961--967},
  year    = {2025},
  doi     = {10.1364/OPTICA.560275}
}

@article{Volkovich2020,
  author  = {Volkovich, S. and Shwartz, S.},
  title   = {Subattosecond x-ray {Hong--Ou--Mandel} metrology},
  journal = {Opt. Lett.},
  volume  = {45},
  number  = {10},
  pages   = {2728--2731},
  year    = {2020},
  doi     = {10.1364/OL.382044}
}

@article{PowersDurbin2026,
  title = {Hong-Ou-Mandel two-photon x-ray states},
  author = {Powers, Liam T. and Durbin, Stephen M.},
  journal = {Phys. Rev. Res.},
  volume = {8},
  issue = {3},
  pages = {L032042},
  numpages = {5},
  year = {2026},
  month = {Sep},
  publisher = {American Physical Society},
  doi = {10.1103/p759-6v4q},
  url = {https://link.aps.org/doi/10.1103/p759-6v4q}
}

@article{Haldane1983PLA,
  author  = {Haldane, F. D. M.},
  title   = {Continuum dynamics of the 1-{D} {Heisenberg} antiferromagnet: Identification with the {O(3)} nonlinear sigma model},
  journal = {Phys. Lett. A},
  volume  = {93},
  pages   = {464},
  year    = {1983},
  doi     = {10.1016/0375-9601(83)90631-X}
}

@article{Haldane1983PRL,
  author  = {Haldane, F. D. M.},
  title   = {Nonlinear field theory of large-spin {Heisenberg} antiferromagnets: Semiclassically quantized solitons of the one-dimensional easy-axis {N}{\'e}el state},
  journal = {Phys. Rev. Lett.},
  volume  = {50},
  pages   = {1153},
  year    = {1983},
  doi     = {10.1103/PhysRevLett.50.1153}
}

@article{AKLT1987,
  author  = {Affleck, Ian and Kennedy, Tom and Lieb, Elliott H. and Tasaki, Hal},
  title   = {Rigorous results on valence-bond ground states in antiferromagnets},
  journal = {Phys. Rev. Lett.},
  volume  = {59},
  pages   = {799},
  year    = {1987},
  doi     = {10.1103/PhysRevLett.59.799}
}

@article{AKLT1988,
  author  = {Affleck, Ian and Kennedy, Tom and Lieb, Elliott H. and Tasaki, Hal},
  title   = {Valence bond ground states in isotropic quantum antiferromagnets},
  journal = {Commun. Math. Phys.},
  volume  = {115},
  pages   = {477},
  year    = {1988},
  doi     = {10.1007/BF01218021}
}

@article{Renard1987,
  author  = {Renard, J. P. and Verdaguer, M. and Regnault, L. P. and Erkelens, W. A. C. and Rossat-Mignod, J. and Stirling, W. G.},
  title   = {Presumption for a quantum energy gap in the quasi-one-dimensional {$S=1$} {Heisenberg} antiferromagnet {NENP}},
  journal = {Europhys. Lett.},
  volume  = {3},
  pages   = {945},
  year    = {1987},
  doi     = {10.1209/0295-5075/3/8/013}
}

@article{Hagiwara1990,
  author  = {Hagiwara, M. and Katsumata, K. and Affleck, Ian and Halperin, B. I. and Renard, J. P.},
  title   = {Observation of {$S=1/2$} degrees of freedom in an {$S=1$} linear-chain {Heisenberg} antiferromagnet},
  journal = {Phys. Rev. Lett.},
  volume  = {65},
  pages   = {3181},
  year    = {1990},
  doi     = {10.1103/PhysRevLett.65.3181}
}

@article{NENP2025,
  title = {Properties of an organic model $S=1$ Haldane chain system},
  author = {Jakovac, Ivan and Cvitani{\ifmmode \acute{c}\else \'{c}\fi{}}, Ton{\ifmmode \check{c}\else \v{c}\fi{}}i and Ar{\ifmmode \check{c}\else \v{c}\fi{}}on, Denis and Herak, Mirta and Cin{\ifmmode \check{c}\else \v{c}\fi{}}i{\ifmmode \acute{c}\else \'{c}\fi{}}, Dominik and Topi{\ifmmode \acute{c}\else \'{c}\fi{}}, Nea Baus and Hosokoshi, Yuko and Ono, Toshio and Iwashita, Ken and Hayashi, Nobuyuki and Amaya, Naoki and Matsuo, Akira and Kindo, Koichi and Lon{\ifmmode \check{c}\else \v{c}\fi{}}ari{\ifmmode \acute{c}\else \'{c}\fi{}}, Ivor and Horvati{\ifmmode \acute{c}\else \'{c}\fi{}}, Mladen and Takigawa, Masashi and Grbi{\ifmmode \acute{c}\else \'{c}\fi{}}, Mihael S.},
  journal = {Phys. Rev. B},
  volume = {111},
  issue = {6},
  pages = {064407},
  numpages = {13},
  year = {2025},
  month = {Feb},
  publisher = {American Physical Society},
  doi = {10.1103/PhysRevB.111.064407},
  url = {https://link.aps.org/doi/10.1103/PhysRevB.111.064407}
}

@article{Y2BaNiO5,
title = {The compound Y2BaNiO5: A new example of a haldane gap in A S = 1 magnetic chain},
journal = {Solid State Communications},
volume = {86},
number = {7},
pages = {409-412},
year = {1993},
issn = {0038-1098},
doi = {https://doi.org/10.1016/0038-1098(93)90455-V},
url = {https://www.sciencedirect.com/science/article/pii/003810989390455V},
author = {J. Darriet and L.P. Regnault}
}

@article{Horodecki2009,
  author = {Horodecki, Ryszard and Horodecki, Pawe{\l} and Horodecki, Micha{\l} and Horodecki, Karol},
  title = {Quantum entanglement},
  journal = {Reviews of Modern Physics}, volume = {81}, pages = {865--942}, year = {2009},
  doi = {10.1103/RevModPhys.81.865}
}

@article{Amico2008,
  author = {Amico, Luigi and Fazio, Rosario and Osterloh, Andreas and Vedral, Vlatko},
  title = {Entanglement in many-body systems},
  journal = {Reviews of Modern Physics}, volume = {80}, pages = {517--576}, year = {2008},
  doi = {10.1103/RevModPhys.80.517}
}

@article{Laflorencie2016,
  author = {Laflorencie, Nicolas},
  title = {Quantum entanglement in condensed matter systems},
  journal = {Physics Reports}, volume = {646}, pages = {1--59}, year = {2016},
  doi = {10.1016/j.physrep.2016.06.008}
}

@article{Islam2015,
  author = {Islam, Rajibul and Ma, Ruichao and Preiss, Philipp M. and Tai, M. Eric and Lukin, Alexander and Rispoli, Matthew and Greiner, Markus},
  title = {Measuring entanglement entropy in a quantum many-body system},
  journal = {Nature}, volume = {528}, pages = {77--83}, year = {2015},
  doi = {10.1038/nature15750}
}

@article{Brydges2019,
  author = {Brydges, Tiff and Elben, Andreas and Jurcevic, Petar and Vermersch, Beno{\^i}t and Maier, Christine and Lanyon, Ben P. and Zoller, Peter and Blatt, Rainer and Roos, Christian F.},
  title = {Probing {R\'enyi} entanglement entropy via randomized measurements},
  journal = {Science}, volume = {364}, pages = {260--263}, year = {2019},
  doi = {10.1126/science.aau4963}
}

@article{Hauke2016,
  author = {Hauke, Philipp and Heyl, Markus and Tagliacozzo, Luca and Zoller, Peter},
  title = {Measuring multipartite entanglement through dynamic susceptibilities},
  journal = {Nature Physics}, volume = {12}, pages = {778--782}, year = {2016},
  doi = {10.1038/nphys3700}
}

@article{Scheie2021,
  author = {Scheie, A. and Laurell, Pontus and Samarakoon, A. M. and Lake, B. and Nagler, S. E. and Granroth, G. E. and Okamoto, S. and Alvarez, G. and Tennant, D. A.},
  title = {Witnessing entanglement in quantum magnets using neutron scattering},
  journal = {Physical Review B}, volume = {103}, pages = {224434}, year = {2021},
  doi = {10.1103/PhysRevB.103.224434}
}

@article{Laurell2021,
  author = {Laurell, Pontus and Scheie, Allen and Mukherjee, Chiron J. and Koza, Michael M. and Enderle, Mechtild and Tylczynski, Zbigniew and Okamoto, Satoshi and Coldea, Radu and Tennant, D. Alan and Alvarez, Gonzalo},
  title = {Quantifying and Controlling Entanglement in the Quantum Magnet {Cs$_2$CoCl$_4$}},
  journal = {Physical Review Letters}, volume = {127}, pages = {037201}, year = {2021},
  doi = {10.1103/PhysRevLett.127.037201}
}

@article{Hales2023,
  author = {Hales, Jordyn and Bajpai, Utkarsh and Liu, Tongtong and Baykusheva, Denitsa R. and Li, Mingda and Mitrano, Matteo and Wang, Yao},
  title = {Witnessing light-driven entanglement using time-resolved resonant inelastic {X}-ray scattering},
  journal = {Nature Communications}, volume = {14}, pages = {3512}, year = {2023},
  doi = {10.1038/s41467-023-38540-3}
}

@misc{Ren2024,
  author = {Ren, Tianhao and Shen, Yao and TenHuisen, Sophia F. R. and Sears, Jennifer and He, Wei and Upton, Mary H. and Casa, Diego and Becker, Petra and Mitrano, Matteo and Dean, Mark P. M. and Konik, Robert M.},
  title = {Witnessing Quantum Entanglement Using Resonant Inelastic {X}-ray Scattering},
  year = {2024}, eprint = {2404.05850}, archivePrefix = {arXiv},
  primaryClass = {cond-mat.str-el}, doi = {10.48550/arXiv.2404.05850},
  note = {arXiv preprint}
}

@misc{TenHuisen2026,
  author = {TenHuisen, Sophia F. R. and Shen, Z. and Bhartiya, V. and Menon, V. and Sharma, P. and Padma, H. and Guan, Z. and He, W. and Lajer, M. K. and Li, J. and Banerjee, D. and Pelliciari, J. and Zaliznyak, I. A. and Gu, G. D. and Lukin, M. D. and Johnston, S. and Dean, M. P. M. and Bisogni, V. and Wang, Y. and Mitrano, M.},
  title = {Observation of multipartite spin entanglement in a cuprate chain},
  year = {2026}, eprint = {2609.08134}, archivePrefix = {arXiv},
  primaryClass = {cond-mat.str-el}, doi = {10.48550/arXiv.2609.08134},
  note = {arXiv preprint}
}

@article{Shen2020,
  author = {Shen, J. and Kuhn, S. J. and Dalgliesh, R. M. and de Haan, V. O. and Geerits, N. and Irfan, A. A. M. and Li, F. and Lu, S. and Parnell, S. R. and Plomp, J. and van Well, A. A. and Washington, A. and Baxter, D. V. and Ortiz, G. and Snow, W. M. and Pynn, R.},
  title = {Unveiling contextual realities by microscopically entangling a neutron},
  journal = {Nature Communications}, volume = {11}, pages = {930}, year = {2020},
  doi = {10.1038/s41467-020-14741-y}
}

@article{Irfan2021,
  author = {Irfan, Abu Ashik Md and Blackstone, Patrick and Pynn, Roger and Ortiz, Gerardo},
  title = {Quantum entangled-probe scattering theory},
  journal = {New Journal of Physics}, volume = {23}, pages = {083022}, year = {2021},
  doi = {10.1088/1367-2630/ac12e0}
}

@article{Cabrillo1999,
  author = {Cabrillo, C. and Cirac, J. I. and Garc{\'i}a-Fern{\'a}ndez, P. and Zoller, P.},
  title = {Creation of entangled states of distant atoms by interference},
  journal = {Physical Review A}, volume = {59}, pages = {1025--1033}, year = {1999},
  doi = {10.1103/PhysRevA.59.1025}
}

@article{DLCZ2001,
  author = {Duan, L.-M. and Lukin, M. D. and Cirac, J. I. and Zoller, P.},
  title = {Long-distance quantum communication with atomic ensembles and linear optics},
  journal = {Nature}, volume = {414}, pages = {413--418}, year = {2001},
  doi = {10.1038/35106500}
}

@article{Bernien2013,
  author = {Bernien, H. and Hensen, B. and Pfaff, W. and Koolstra, G. and Blok, M. S. and Robledo, L. and Taminiau, T. H. and Markham, M. and Twitchen, D. J. and Childress, L. and Hanson, R.},
  title = {Heralded entanglement between solid-state qubits separated by three metres},
  journal = {Nature}, volume = {497}, pages = {86--90}, year = {2013},
  doi = {10.1038/nature12016}
}

@article{Delteil2016,
  author = {Delteil, Aymeric and Sun, Zhe and Gao, Wei-bo and Togan, Emre and Faelt, Stefan and Imamoglu, Ata{\c c}},
  title = {Generation of heralded entanglement between distant hole spins},
  journal = {Nature Physics}, volume = {12}, pages = {218--223}, year = {2016},
  doi = {10.1038/nphys3605}
}

@article{Stockill2017,
  author = {Stockill, R. and Stanley, M. J. and Huthmacher, L. and Clarke, E. and Hugues, M. and Miller, A. J. and Matthiesen, C. and Le Gall, C. and Atat{\"u}re, M.},
  title = {Phase-Tuned Entangled State Generation between Distant Spin Qubits},
  journal = {Physical Review Letters}, volume = {119}, pages = {010503}, year = {2017},
  doi = {10.1103/PhysRevLett.119.010503}
}

@article{Lee2011,
  author = {Lee, K. C. and Sprague, M. R. and Sussman, B. J. and Nunn, J. and Langford, N. K. and Jin, X.-M. and Champion, T. and Michelberger, P. and Reim, K. F. and England, D. and Jaksch, D. and Walmsley, I. A.},
  title = {Entangling macroscopic diamonds at room temperature},
  journal = {Science}, volume = {334}, pages = {1253--1256}, year = {2011},
  doi = {10.1126/science.1211914}
}

@article{hepting2020electronic,
	Author = {Hepting, M. and Li, D. and Jia, C. J. and Lu, H. and Paris, E. and Tseng, Y. and Feng, X. and Osada, M. and Been, E. and Hikita, Y. and Chuang, Y. -D. and Hussain, Z. and Zhou, K. J. and Nag, A. and Garcia-Fernandez, M. and Rossi, M. and Huang, H. Y. and Huang, D. J. and Shen, Z. X. and Schmitt, T. and Hwang, H. Y. and Moritz, B. and Zaanen, J. and Devereaux, T. P. and Lee, W. S.},
	Da = {2020/04/01},
	Doi = {10.1038/s41563-019-0585-z},
	Id = {Hepting2020},
	Isbn = {1476-4660},
	Journal = {Nature Materials},
	Number = {4},
	Pages = {381--385},
	Title = {Electronic structure of the parent compound of superconducting infinite-layer nickelates},
	Ty = {JOUR},
	Url = {https://doi.org/10.1038/s41563-019-0585-z},
	Volume = {19},
	Year = {2020}}

@article{jia2014persistent,
	Author = {Jia, C. J. and Nowadnick, E. A. and Wohlfeld, K. and Kung, Y. F. and Chen, C. -C. and Johnston, S. and Tohyama, T. and Moritz, B. and Devereaux, T. P.},
	Da = {2014/02/28},
	Doi = {10.1038/ncomms4314},
	Id = {Jia2014},
	Isbn = {2041-1723},
	Journal = {Nature Communications},
	Number = {1},
	Pages = {3314},
	Title = {Persistent spin excitations in doped antiferromagnets revealed by resonant inelastic light scattering},
	Ty = {JOUR},
	Url = {https://doi.org/10.1038/ncomms4314},
	Volume = {5},
	Year = {2014}}

@article{devereaux2016directly,
  title = {Directly Characterizing the Relative Strength and Momentum Dependence of Electron-Phonon Coupling Using Resonant Inelastic X-Ray Scattering},
  author = {Devereaux, T. P. and Shvaika, A. M. and Wu, K. and Wohlfeld, K. and Jia, C. J. and Wang, Y. and Moritz, B. and Chaix, L. and Lee, W.-S. and Shen, Z.-X. and Ghiringhelli, G. and Braicovich, L.},
  journal = {Phys. Rev. X},
  volume = {6},
  issue = {4},
  pages = {041019},
  numpages = {12},
  year = {2016},
  month = {Oct},
  publisher = {American Physical Society},
  doi = {10.1103/PhysRevX.6.041019},
  url = {https://link.aps.org/doi/10.1103/PhysRevX.6.041019}
}

@article{schlappa2012spin,
	Author = {Schlappa, J. and Wohlfeld, K. and Zhou, K. J. and Mourigal, M. and Haverkort, M. W. and Strocov, V. N. and Hozoi, L. and Monney, C. and Nishimoto, S. and Singh, S. and Revcolevschi, A. and Caux, J. -S. and Patthey, L. and R{\o}nnow, H. M. and van den Brink, J. and Schmitt, T.},
	Da = {2012/05/01},
	Doi = {10.1038/nature10974},
	Id = {Schlappa2012},
	Isbn = {1476-4687},
	Journal = {Nature},
	Number = {7396},
	Pages = {82--85},
	Title = {Spin--orbital separation in the quasi-one-dimensional Mott insulator Sr2CuO3},
	Ty = {JOUR},
	Url = {https://doi.org/10.1038/nature10974},
	Volume = {485},
	Year = {2012}}

@article{rossi2019experimental,
  title = {Experimental Determination of Momentum-Resolved Electron-Phonon Coupling},
  author = {Rossi, Matteo and Arpaia, Riccardo and Fumagalli, Roberto and Moretti Sala, Marco and Betto, Davide and Kummer, Kurt and De Luca, Gabriella M. and van den Brink, Jeroen and Salluzzo, Marco and Brookes, Nicholas B. and Braicovich, Lucio and Ghiringhelli, Giacomo},
  journal = {Phys. Rev. Lett.},
  volume = {123},
  issue = {2},
  pages = {027001},
  numpages = {6},
  year = {2019},
  month = {Jul},
  publisher = {American Physical Society},
  doi = {10.1103/PhysRevLett.123.027001},
  url = {https://link.aps.org/doi/10.1103/PhysRevLett.123.027001}
}

@article{Shih2001,
  title = {Two-Photon Diffraction and Quantum Lithography},
  author = {D'Angelo, Milena and Chekhova, Maria V. and Shih, Yanhua},
  journal = {Phys. Rev. Lett.},
  volume = {87},
  issue = {1},
  pages = {013602},
  numpages = {4},
  year = {2001},
  month = {Jun},
  publisher = {American Physical Society},
  doi = {10.1103/PhysRevLett.87.013602},
  url = {https://link.aps.org/doi/10.1103/PhysRevLett.87.013602}
}

\end{document}